\documentclass[article, twocolumn, aps, prl, groupedaddress, showpacs, superscriptaddress, nofootinbib]{revtex4-2}

\usepackage{amsmath}
\usepackage{amssymb}
\usepackage{txfonts}
\usepackage{gensymb}
\usepackage{graphicx,bm,units,yfonts}
\usepackage[table]{xcolor} 
\usepackage{hyperref} 
\usepackage{cancel} 
\usepackage[T1]{fontenc} 
\usepackage{lmodern}
\usepackage{changes}
\usepackage{romannum} 
\usepackage{array}
\usepackage{braket}

\begin{document}

\title{Observation of Phase Controllable Majorana-like Bound States in Metamaterial-based Kitaev Chain Analogues}

\author{Kai Qian}
\altaffiliation[Present address: ]
{Department of Mechanical Aerospace and Engineering, University of California San Diego, La Jolla, California 92093, USA}
\affiliation{Department of Physics, New Jersey Institute of Technology, Newark, New Jersey, 07102, USA}

\author{David J. Apigo}
\affiliation{Department of Physics, New Jersey Institute of Technology, Newark, New Jersey, 07102, USA}

\author{Karmela Padavi{\'c}}
\altaffiliation[Present address: ]
                     {Department of Science, Bard High School Early College, New York, New York 10002, USA}
\affiliation{Department of Physics, University of Illinois at Urbana-Champaign, Urbana, Illinois, 61801, USA}

\author{Keun~Hyuk~Ahn}
\email{kenahn@njit.edu}
\affiliation{Department of Physics, New Jersey Institute of Technology, Newark, New Jersey, 07102, USA}

\author{Smitha Vishveshwara}
\affiliation{Department of Physics, University of Illinois at Urbana-Champaign, Urbana, Illinois, 61801, USA}

\author{Camelia Prodan}
\email{cprodan@njit.edu} 
\affiliation{Department of Physics, New Jersey Institute of Technology, Newark, New Jersey, 07102, USA}

\begin{abstract}
We experimentally demonstrate that Majorana-like bound states (MLBSs) can occur in quasi-one-dimensional metamaterials, analogous to Majorana zero modes (MZM) in the Kitaev chain.
In a mechanical spinner ladder system, we observe a topological phase transition and spectral-gap-protected edge MLBSs. 
We characterize the decaying and oscillatory nature of these MLBS pairs and their phase-dependent hybridization. 
It is shown that the hybridization can be tuned to yield the analogue of parity switching in MZMs, a key element of topological qubits.
We find strong agreements with theory.
\end{abstract}

\maketitle

\setlength{\abovedisplayskip}{3pt}
\setlength{\belowdisplayskip}{3pt}

In the past decade, the rich cross-fertilization of ideas 
between the studies of  electronic and mechanical systems has led 
to the discovery of exciting new states of matter, 
including physical behaviors driven by topology~\cite{Prodan09,Kane14,Chen14,Huber16,Chen16,Gao19}.
For example, a Chern insulator with one-way edge modes was demonstrated using an array of gyroscopes~\cite{Nash15, Wang15}.
Among electronic materials, topological superconductors
have gained prominence as candidates for hosting Majorana zero 
modes (MZMs), which are potential 
building blocks of fault tolerant quantum 
computing~\cite{Majorana37,Kitaev03,Nayak08,Wilczek09,Lutchyn10,Oreg10,Alicea12, Hosur11, Leijnse12, Teo13,  Nadj-Perge14, Sato16}. 
The Kitaev chain~\cite{Kitaev01}, consisting of electrons hopping on a lattice subject to $p$-wave pairing, 
provides an excellent prototype for realizing MZMs as topologically protected edge states~\cite{Chubb16, Lian18, Aguado20, Tutschku20}.  
Remarkably, these modes have remained elusive in solid state systems 
despite tremendous efforts~\cite{Hughes11,Mourik12,Rokhinson12, Das12,Albrecht16,Deng16,Lutchyn18,Wang20}. 
Nevertheless, MZMs have continued to inspire experimental 
realizations of analogous modes in other systems~\cite{McDonald18,Vishveshwara21},
including metamaterials. 
Even key traits for error-free MZM qubits, like non-Abelian braiding, have recently been proposed in classical metamaterials systems~\cite{Barlas20}.

\begin{figure*}
\center
\includegraphics[width=1\linewidth]{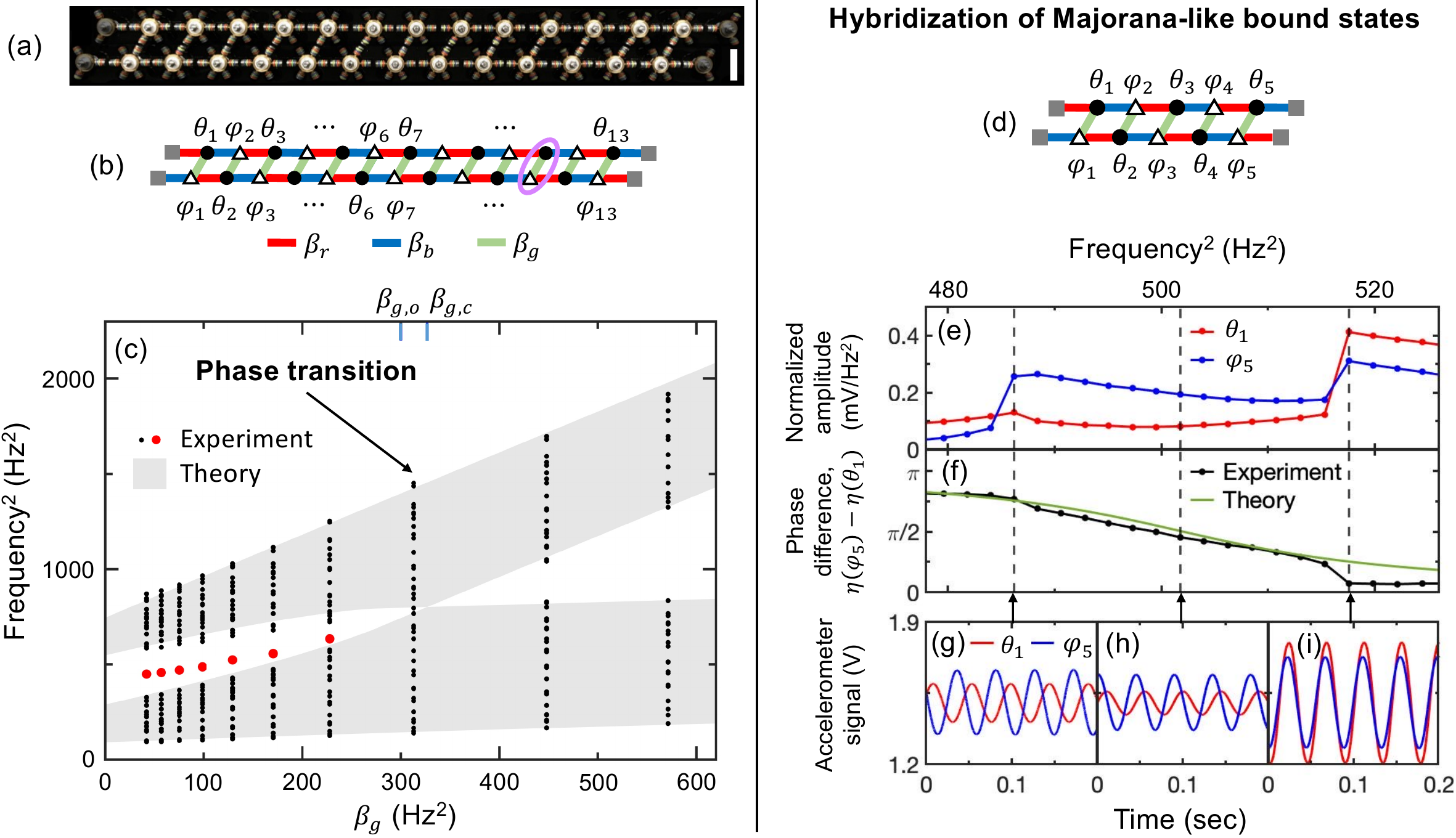}
\caption{\small 
Phase transition for a long ladder system (left) and phase-controllable hybridized MLBSs for a short ladder system (right).
(a) Picture of the experimental setup for a system of length $N$=13. Rotatable spinners and magnetically coupled arms are highlighted and fixed spinners and arms without magnets are shaded. The scale on the right  is 5.0 cm. 
(b) Schematic diagram for the SSH ladder system pictured in (a). The rotatable spinners are represented by black circles ($\theta_n$) and open triangles ($\varphi_n$), and the fixed spinners by gray squares. The purple ellipse shows the unit cell. The red and blue lines denote the {\it intrachain} coupling $\beta_r=230~\text{Hz}^2$
and $\beta_b=100~\text{Hz}^2$, respectively, 
and the green line the {\it interchain} coupling $\beta_g$. 
(c) Topological phase transition for the long $N$=13 system 
signaled by the emergence of mid-gap MLBSs. 
The measured frequency squared ($f^2$) for normal mode 
versus $\beta_g$ is shown for various interchain coupling $\beta_g$.  
The red dots correspond to the mid-gap MLBSs, 
while the black dots bulk states. 
The gray areas represent the theoretical bulk bands 
in the limit of the infinitely long system.  
The two critical interchain couplings, 
$\beta_{g,c}=330~\text{Hz}^2$ for the phase transition
and $\beta_{g,o}=300~\text{Hz}^2$ for oscillatory MLBSs, 
are marked on the top axis. 
(d) Schematic diagram for the short $N$=5 system. 
(e) Spectra obtained by actuating $\theta_1$ spinner 
and measuring at $\theta_1$ and $\varphi_5$ spinners, 
shown in red and blue dots, respectively,
within the bulk gap 
for the topological phase with $\beta_g$= 100~Hz$^2$
and $\beta_r$ and $\beta_b$ identical to those for (c).  
A single degenerate MLBS peak in (c) is split into two MLBS peaks 
due to the enhanced hybridization. 
(f) Phase difference between the $\varphi_5$ and $\theta_1$ spinners, 
$\eta(\varphi_5)-\eta(\theta_1)$, versus $f^2$, 
revealing phase controllable MLBSs.
(g), (h) and (i): Three examples of the accelerometer signals, 
proportional to the rotation angles, versus time
for lower MLBS, in-between, and upper MLBS, 
as marked by arrows.
The red and blue lines represent the oscillations of  
the $\theta_1$ and $\varphi_5$ spinners.
(See Supplemental Material~\cite{SupMat} for 
typical spectra, gap versus $\beta_g$, and 
videos for modes.)
}
\label{figure1}
\end{figure*}

In this Letter, we present a mechanical metamaterial ladder system made of magnetically coupled spinners, and demonstrate that it closely parallels several salient features of the Kitaev chain, including distinct presence of Majorana-like bound states (MLBSs). 
In previous work by two of the current authors and collaborators~\cite{Padavic18}, 
it was shown that the phase diagram, mode spectrum, and associated wavefunctions of the Kitaev chain could be mapped to a system of two coupled 
Su-Schrieffer-Heeger (SSH)
chains~\cite{Su79}, characterized by alternating bond strengths schematically shown in Fig.~\ref{figure1}(b). 
A significant benefit of this mapping is 
that the SSH ladder system 
can be realized in a variety of systems whether they be bosonic, fermionic or involve classical metamaterials. 
Although MLBSs in these systems may not be exactly equivalent to MZMs, 
they are topologically protected by spectral gaps. In a coupled split-ring-resonator system, Guo {\it et al.}~\cite{Guo21} observed one such metamaterial realization of MLBSs and a topological phase transition up to finite size effects. 
Here, our work on the spinner system explores a topological phase transition for sufficiently long ladders, thus eliminating size effects, and the properties of the MLBS at the edges that were theoretically predicted~\cite{Padavic18,Kao14,Hegde15,Hegde16} but never fully realized.

In what follows, we demonstrate the remarkable analogues to Kitaev chain features in metamaterials, including spectral characteristics of topological and nontopological phases, oscillatory, decaying MZM wavefunctions, and switching behavior in fermion parity, which forms the basis of MZM qubits.
The Kitaev chain is described by the Hamiltonian~\cite{Kitaev01,Padavic18}, 
\begin{equation} \label{eq:HKitaev}
H_\text{K}=\frac{1}{2i}\sum_{n}^{}(\omega-\Delta)a_{n}b_{n+1}+(\omega+\Delta)a_{n+1}b_{n}+\mu a_nb_n,
\end{equation}
where $a_n$ and $b_n$ are the Majorana operators, 
$\omega$ the nearest neighbor electron hopping amplitude, 
$\Delta$ the $p$-wave superconducting order parameter, 
and $\mu$ the chemical potential.
One of the two topological gapped phases, hosting mid-gap MZMs,
occurs at $|\mu|<2|\omega|$ and $\Delta>0$
and the other at  $|\mu|<2|\omega|$ and $\Delta<0$,
while nontopological gapped phases arise at $|\mu|>2|\omega|$. 

Mechanical systems of magnetically 
coupled spinners~\cite{Apigo18, Qian18, Qian20, Zhu19} 
have recently been used to simulate a number of electronic tight-binding Hamiltonians. The spinner ladder system pictured in Fig.~\ref{figure1}(a) shows the metamaterial analogue of the SSH ladder in Fig.~\ref{figure1}(b).
Here the electron hopping translates to the interspinner magnetic interaction
controlled by the distance between magnets,
and the electronic eigenstate energy maps to
the frequency squared ($f^2$) for the spinner normal modes.
With magnets attached to selected spinner arms,
nearest-neighbors connected by the red, blue, and green lines in Fig.~\ref{figure1}(b)
have attractive interactions,
parameterized by
normalized positive constants $\beta_r$, $\beta_b$, and $\beta_g$, respectively. 
As marked by solid circles and open triangles
in Fig.~\ref{figure1}(b), the system is bipartite,
and the rotations of the spinners are
represented by $\theta_n$ and $\varphi_n$ with $n=1,...,N$ 
within each sublattice
for the system of length $N$. 
One of the spinners is driven externally, 
and attached accelerometers monitor oscillations. 
Details of the setup are provided in Supplemental Material~\cite{SupMat} and Refs.~\cite{Apigo18, Qian18, Qian20}. 
The normalized Lagrangian for the spinner system is $L=T-U_{1}-U_{2}$, 
where $T=\sum_{n=1}^{N}(\dot{\theta}_n^2+\dot{\varphi}_n^2)/8\pi^2$ is the normalized kinetic energy,  
and $U_{1}=-\sum_{n=1}^{N} \alpha (\theta_n^2+\varphi_n^2)/2$
with a positive coefficient $\alpha$
and 
\begin{equation} \label{eq:U2}
U_{2}=-\sum_{n}^{} \beta_b \theta_n \varphi_{n+1} + \beta_r \theta_{n+1} \varphi_n+\beta_g \theta_n \varphi_n
\end{equation}
represent the normalized potential energy. 
While the term $U_{1}$ shifts the spectrum by a constant, 
the term $U_{2}$ has the form identical to 
$H_\text{K}$ in Eq.~(\ref{eq:HKitaev}),
depicted by the mapping 
shown in Table~\ref{table1}. This mapping allows us to explore the analogy between the metamaterials and electronic systems through direct excitations of spinners, as shown in videos in Supplemental Material~\cite{SupMat}.

\begin{table}
\centering
\begin{tabular}{ c | c | c | c | c | c } 
 \hline \hline
 SSH spinner ladder system & $\theta_n$ & $\varphi_n$ & $\beta_b$ & $\beta_r$ & $\beta_g$  \\ [0.5ex] 
 \hline
 Electronic Kitaev chain & $a_n$ & $b_n$ & $\omega-\Delta$ & $\omega+\Delta$ & $\mu$ \\ [0.5ex] 
 \hline 
\end{tabular}
\caption{
Mapping between the parameters of the SSH spinner ladder system [Eq.~(\ref{eq:U2})]
and the electronic Kitaev chain [Eq.~(\ref{eq:HKitaev})].
For the spinner system,
the $\theta_n$ and $\varphi_n$ represent the rotation angles of the spinners, 
the $\beta_b$ and $\beta_r$ the {\it intrachain} interactions,
and the $\beta_g$ the {\it interchain} interaction,
as shown in Figs.~\ref{figure1}(b) and \ref{figure1}(d).
For the Kitaev chain,
$a_n$ and $b_n$ are the Majorana operators,
$\omega$ the nearest neighbor electron hopping amplitude, 
$\Delta$ the $p$-wave superconducting order parameter, 
and $\mu$ the chemical potential~\cite{Padavic18}.
}
\label{table1}
\end{table}

In long ladders, we observe a clear phase transition between nontopological and
topological phases characterized by mid-gap MLBS as a function of the interaction between SSH chains,
which is analogous to the chemical potential in the Kitaev chain.
Figure~\ref{figure1}(c) shows 
$f^2$ for normal modes
versus the interchain coupling $\beta_g$ 
for the $N$=13 spinner systems 
[see Figs.~\ref{figure1}(a) and ~\ref{figure1}(b)]
with intrachain couplings $\beta_r=230~\text{Hz}^2$ and $\beta_b=100~\text{Hz}^2$. 
Bulk modes, shown as black dots, are identified 
from the resonant 
oscillations of spinners near the center of the system. 
In contrast, the red dot in the spectrum corresponds to the MLBS at one of the system's ends, and is
prominently identified from measurement of the $\theta_1$ spinner. 
We expect another  MLBS mode
at the other end at the same frequency.
(See Supplemental Material for typical spectra~\cite{SupMat}.) 
Starting from large interchain coupling $\beta_g$, 
the gap in the bulk spectra rapidly narrows as $\beta_g$ decreases,
and closes when $\beta_g\approx$ 310 Hz$^2$, 
consistent with the theoretical phase boundary $\beta_{g,c}=\beta_r+\beta_b=330~\text{Hz}^2$ 
(see the top axis). 
As $\beta_g$ decreases further, 
the gap reopens but with a distinct mid-gap mode, 
which marks the topological phase transition. 
Bulk bands fall within the theoretically predicted range
for the infinitely long ladder system~\cite{Padavic18} 
shown in gray. 
We conclude that the $N$=13 spinner system realizes
the long ladder limit, 
thus expanding on the results 
in Ref.~\cite{Guo21}, where finite size effects presented more of a limitation. 
(See Supplemental Material for
gap versus $\beta_g$ and theoretical ranges for bulk bands~\cite{SupMat}.) 

Realizing the short topological systems, 
we find that MLBSs from the two ends overlap, 
and therefore 
the degenerate end modes of the long limit hybridize
and split in frequency, 
in close analogy to hybridization of MZM wavefunctions.
To first order, the split modes correspond to symmetric and antisymmetric combinations
of the MLBS up to a tunable relative phase. 
For the infinitely long topological system, theory predicts 
that MLBSs reside on the $\theta_n$-sublattice at the left edge and 
on the $\varphi_n$-sublattice at the right edge if $\beta_r > \beta_b$ (as in all cases in this Letter). 
For the short topological system with $N$=5, schematically shown in Fig.~\ref{figure1}(d), 
the hybridization and splitting is evident in Fig.~\ref{figure1}(e),
where the normalized amplitudes 
measured at the $\theta_1$ [$\varphi_5$] spinner 
are plotted with respect to $f^2$ within the gap
in red [blue] dots.
To reveal the impact of the significant overlap of the two peaks,
we measure the angles, $\theta_1$ and $\varphi_5$, 
versus time at three frequencies indicated 
by the dashed lines in Fig.~\ref{figure1}(e), that is,
the lower and upper MLBS frequencies, 
and a frequency in between, 
as shown in Figs.~\ref{figure1}(g), \ref{figure1}(i), and \ref{figure1}(h), respectively. 
The phase differences  
between $\varphi_5$ and $\theta_1$, $\eta(\varphi_5)-\eta(\theta_1)$,
are $0.77\pi$ and $0.07\pi$ 
for the lower and upper MLBSs, respectively. 
In being 
close to $\pi$ or 0, they mimic
the theoretical results for an ideal system without damping,
that is, odd or even symmetry.
Similarly, for a series of frequencies in the gap,
the relative phase $\eta(\varphi_5)-\eta(\theta_1)$
is plotted as dots in Fig.~\ref{figure1}(f). 
It shows a continuous change of the phase difference,
in agreement with theoretical results for a damped system shown in green line. 
Here, we emphasize that the tunability of the relative phase
in our system
may have future applications in mechanical memories~\cite{Hasan21}.

\begin{figure}
\center
\includegraphics[width=\linewidth]{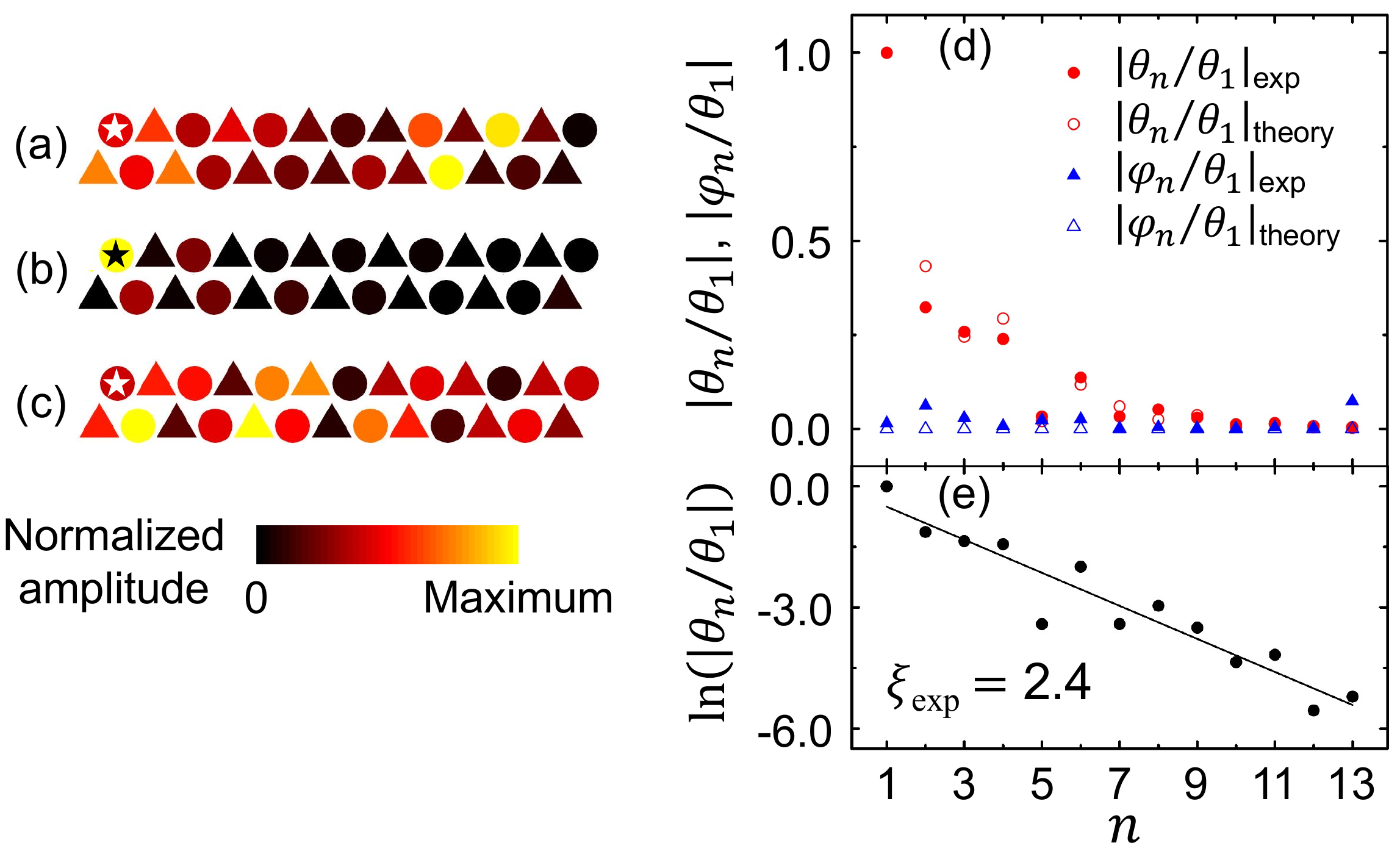}
\caption{\small Localization of the
MLBS for a system with  $N=13$ and $\beta_g$ = 100 Hz$^2$, a case in Fig.~\ref{figure1}(c).
(a)-(c) Rotational oscillation amplitudes of spinners measured for (a) an upper bulk mode,
(b) a mid-gap MLBS, 
and (c) a lower bulk mode
(See Supplemental Material~\cite{SupMat} for videos). 
The color represents the rotational oscillation amplitude of each spinner. 
The stars mark the actuated spinners. 
(d) Normalized rotational oscillation amplitudes, ${|}\theta_n{/}\theta_1{|}$ (red circles) and ${|}\varphi_n{/}\theta_1{|}$ (blue triangles), 
versus $n$. 
The solid and open symbols represent the experimental and theoretical results. 
(e) Symbols: semilogarithmic plot of $\ln{{|}\theta_n{/}\theta_1{|}}$ versus $n$ for the experiments. 
Line: linear fit giving a localization length $\xi_{\rm exp}=2.4$, 
consistent with the theory.}
\label{figure2} 
\end{figure}

Importantly, MLBSs are not only bipartite and decaying, 
but also characterized by spatial oscillations. 
For instance, for the topological $N$=13 system
with $\beta_g$ = 100 Hz$^2$
in Fig.~\ref{figure1}(c), 
rotational oscillation amplitudes of the spinners
for the MLBS are
displayed in colors in Fig.~\ref{figure2}(b). 
Here, it is revealed that the MLBS resides mostly 
on the $\theta_n$ sites (circles) near the left edge,
unlike typical bulk modes shown 
in Figs.~\ref{figure2}(a) and \ref{figure2}(c)
(See Supplemental Material for videos~\cite{SupMat}.).
The detailed structure of the MLBS is seen in Fig.~\ref{figure2}(d), 
where the normalized amplitude 
for the $\theta_n$ ($|\theta_n/\theta_1|$, red solid circles) 
is large near the left edge,
while amplitude for the $\varphi_n$
($|\varphi_n/\theta_1|$, blue solid triangles)
is negligible.
These results qualitatively agree 
with Ref.~\cite{Guo21}, 
but with an important difference: 
in our system the decaying amplitude {\it oscillates}. 
Moreover, this oscillation rather successfully matches theoretical predictions~\cite{Hegde16} and is the first to realize them. 
For MZMs in the Kitaev chain, theory~\cite{Hegde16} predicts
such oscillatory wavefunctions
within a circle 
in the phase diagram
set by $\mu^2 + (2\Delta)^2 < (2\omega)^2$.
Translated to the spinner system, 
the spatial profile of the MLBS localized 
at the left edge is given by 
$\theta_n=A\exp(-n/\xi_{\rm theory})\cos{(B+2n\pi/\lambda)}$
if $\beta_g < \beta_{g,o}$, 
where 
$\xi_{\rm theory}=2[\ln{(\beta_r/\beta_b)}]^{-1}$ is the localization length,
$\lambda$, $A$, and $B$ are constants, 
and $\beta_{g,o}=2\sqrt{\beta_b\beta_r}$
(see Supplemental Material 
for details~\cite{SupMat}).
For the system in Fig.~\ref{figure2},
$\beta_g$= 100 Hz$^2$ is indeed less 
than $\beta_{g,o}$=300 Hz$^2$ 
[see the top axis in Fig.~\ref{figure1}(c)],
and the theoretical results  
for the $|\theta_n/\theta_1|$ and $|\varphi_n/\theta_1|$
plotted in open symbols 
in Fig.~\ref{figure2}(d)
agree well with the experiments.
Further, a linear fit of $\ln{|\theta_n/\theta_1|}$ versus $n$ 
for the experiments shown in Fig.~\ref{figure2}(e)
results in $\xi_{\rm exp}$=2.4, 
identical to the theoretical value $\xi_{\rm theory}$.
Strong agreement with the theory 
confirms that the spatial oscillation seen here
is not due to disorder,
but reflects the intrinsic nature of the MLBS spatial profiles,
analogous to the oscillatory MZM wavefunctions.

\begin{figure}[t]
\includegraphics[width=0.85\linewidth]{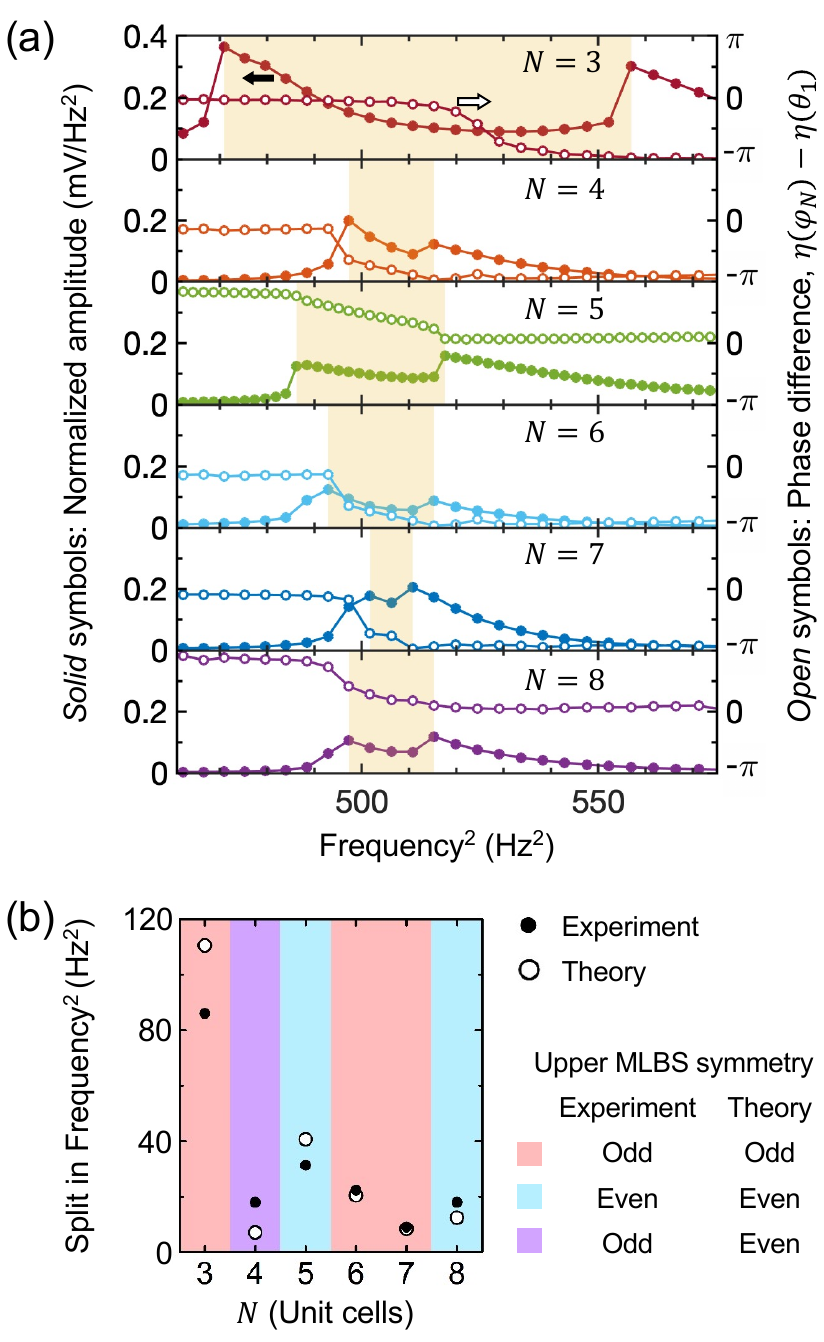}
\caption{\small Split oscillation and phase difference switching of the hybridized MLBSs
as varying the length of the systems.
(a) {\it Solid} symbols: Spectra  of the topological systems for $N=3, 4, ..., 8$,
obtained by actuating $\theta_1$ and measuring at $\varphi_N$ spinners for the frequencies 
within the bulk gap.
The interactions are kept identical to those in Figs.~\ref{figure1}(d)-\ref{figure1}(i).
The widths of the yellow areas represent the splits of the hybridized MLBSs.  
{\it Open} symbols: The corresponding phase differences, $\eta(\varphi_N)-\eta(\theta_1)$. 
(b) Symbols: Split between the hybridized MLBSs versus the length of the system, $N$, 
for the experiments (solid symbols) 
and the theory (open symbols). 
Background colors:
Observed phase difference 
of the upper MLBS is shown 
in blue for close to 0 (even symmetry)
and red for $\pm \pi$ (odd), 
agreeing with the theory.
A disagreement between theory and experiments
occurs for $N$=4 case, shown in purple, 
likely due to disorder in the experimental setup~\cite{Footnote}.
}
\label{figure3}
\end{figure}

The oscillatory behavior of the MLBSs significantly affects spectral features of the system. 
Most importantly, when their hybridization changes its sign, there is a switch between whether the symmetric or the antisymmetric combination corresponds to the lower frequency. 
In the MZM analogue, this behavior results in so-called
 parity switching~\cite{Kao14,Hegde15,Hegde16,Avila20}. 
Here, consequences of the switch are rather profound: the two
end MZMs form a Dirac fermionic state that can be occupied or unoccupied, essentially a topological qubit, thus giving the system odd or even parity.
In other words, the sign of the hybridization determines ground state parity. 
To demonstrate the analogue of this switching in the topological spinner systems, 
we vary the length of the system, $N$=3, 4,..., 8, 
while keeping the interactions identical to
Figs.~\ref{figure1}(d)-\ref{figure1}(i). 
Normalized amplitudes of
the $\varphi_N$ spinner 
(resulting from actuating the $\theta_1$ spinner)
versus $f^2$ within the bulk gap 
are shown as {\it solid} symbols in Fig.~\ref{figure3}(a).
The split of the MLBSs, indicated by
yellow areas in Fig.~\ref{figure3}(a)
and solid circles in Fig.~\ref{figure3}(b),
clearly oscillates as $N$ increases,  
in agreement with theoretical results,
shown as open circles in Fig.~\ref{figure3}(b). 
Behavior of the spinner system analogous to parity switching 
is most evident in
the phase difference between the $\varphi_N$ and $\theta_1$ spinner,
$\eta(\varphi_N)-\eta(\theta_1)$, versus $f^2$
shown as {\it open} symbols in Fig.~\ref{figure3}(a).
The phase difference changes approximately 
from $\pi$ to 0 
as the frequency increases for $N$= 5 and 8, 
but from 0 to $-\pi$ for $N$=3, 4, 6, and 7.
The observed switch in the phase difference
for the upper MLBS 
is displayed in Fig.~\ref{figure3}(b)
using the background color for each $N$,
that is,
blue for $\eta(\varphi_N)-\eta(\theta_1) \approx 0$ (even symmetry)
and red for $\pm\pi$ (odd). 
It agrees with the theory
except for the $N$=4 case 
(purple, odd from the experiments, but even from the theory)
~\cite{Hegde16,Footnote}. 
The phase switching seen here may prove to be a consequential feature when it comes to designing MZM qubits, 
because it strengthens the proposal
that the chemical potential be tuned to 
achieve exactly degenerate MZMs in finite Kitaev chains~\cite{Kao14,Hegde15,Hegde16}.

In conclusion, demonstrating the potentiality of research that connects electronic and metamaterial systems,
we analyzed surprisingly close analogues of MZMs in Kitaev chains with MLBSs in spinner ladders. 
While the degrees of freedom differ, our in-depth realization provides a well-controlled prototype for benchmarking any MZMs 
and Majorana-based qubits that may be discovered in the future~\cite{Sau20}. 
Further, our studies signal promise for next-generation metamaterial applications, such as topological braiding
of mid-gap states in the case of stacked SSH ladders~\cite{Barlas20}. From the metamaterials perspective,
states like MLBSs that are protected by spectral gaps offer wide possibilities, such as mechanical memory and device applications.

K.Q. and C.P. acknowledge support from the W. M. Keck Foundation. C. P. acknowledges support from National Science Foundation (NSF Award 2131759). K.Q., K.H.A. and C.P acknowledge support of NJIT Faculty Seed Grant.
S.V. acknowledges support of the National Science Foundation and the Quantum Leap Challenge Institute for Hybrid Quantum Architectures and Networks (NSF Award 2016136).


\begin{thebibliography}{}
\bibitem{Footnote} 
The difference between theory and experiments for $N$=4 system is likely originated from disorders 
present in the experimental setup.
For example, the two theoretically predicted MLBS peaks 
are close enough 
that they could be switched in the experiments 
due to disorders. 
Possible source of the disorders includes 
the moments of inertia of the attached accelerometers.
\end{thebibliography}
\end{document}